\documentclass[aps,pra,twocolumn,showpacs,superscriptaddress,floatfix]{revtex4}
\usepackage[utf8]{inputenc}
\usepackage{amsmath}
\usepackage{amssymb}
\usepackage{amsthm}
\usepackage{mathtools}
\usepackage{graphicx}
\usepackage{xcolor}
\usepackage{hyperref,lipsum,ulem}
\usepackage{dpackage}

\hypersetup{
  colorlinks   = true, 
  urlcolor     = red, 
  linkcolor    = blue, 
  citecolor   = violet 
} 

\begin{document}

\title{Degradation of Entanglement in Markovian Noise}
\author{Dario Gatto}
\affiliation{Dipartimento di Fisica, Universit\'{a} di Pisa, Largo Bruno Pontecorvo 3, 56126 Pisa, Italy}

\author{Antonella De Pasquale}
\affiliation{Dipartimento di Fisica e Astronomia, Universit\'{a} di Firenze, I-50019, Sesto Fiorentino (FI), Italy}
\affiliation{INFN, Sezione di Firenze, I-50019, Sesto Fiorentino (FI), Italy}
\affiliation{NEST, Scuola Normale Superiore and Istituto Nanoscienze-CNR, Piazza dei Cavalieri 7, I-56126 Pisa, Italy}

\author{Vittorio Giovannetti}
\affiliation{NEST, Scuola Normale Superiore and Istituto Nanoscienze-CNR, Piazza dei Cavalieri 7, I-56126 Pisa, Italy}


\begin{abstract}
The entanglement survival time is defined as the maximum time a system which  
is evolving under the action of local Markovian, homogenous in time noise,
is capable to preserve the entanglement it had at the beginning of the temporal evolution. In this paper we study how this quantity is affected by 
the interplay between the coherent preserving and dissipative contributions of the corresponding  dynamical generator. We report the presence of a
 counterintuitive,  non-monotonic behaviour in such functional, capable of 
inducing  sudden death of entanglement in models  which, in the absence 
of unitary driving are capable to sustain entanglement for arbitrarily long times. 
\end{abstract}

\maketitle

\section{Introduction}
Entanglement is a fundamental, yet extremely fragile resource of quantum 
information processing~\cite{entdeath}. Preventing its degradation  
is a fundamental step in the development of  quantum technology. 
 Starting from the seminal work on quantum error correction~\cite{Preskill}, decoherence-free subspaces~\cite{Lidar}, and dynamical decoupling~\cite{viola}
a number of methods have been proposed to provide partial protection against such detrimental effect.
Most of these approaches typically work under the paradigm of  mitigating the environmental noise by properly 
intertwining  the dynamics it induces  with external controls. Moreover,  such controls usually correspond to Hamiltonian 
corrections.The basic idea is to fight dissipative and decoherence mechanisms through the action of driving forces that drag the system
in  regions of the Hilbert space where the former are no so effective. 
Interestingly enough such external forces do 
not necessarily need to be coherent preserving: indeed, while typically 
summing noise sources tends to add up speed at which entanglement get 
lost~\cite{mix}, it may occur that by properly alternating their actions the 
entanglement survival time can 
be increased~\cite{CuT}.
Similarly, it is clear that not always coherence-preserving controls help in contrasting the noise:  a not carefully designed Hamiltonian driving  might amplify the dissipation induced
by the environment.
Motivated  by these observations, in the present paper we study the maximum entanglement survival time $\tau_{ent}$  for a system evolving under the action of a local 
Markovian, time-homogenous noise~\cite{BOOK}. 
In the general  formalism established by  Gorini, Kossakowski, Sudarshan, and Lindblad~\cite{gorini76,lindblad76} these  models are fully described by assigning a dynamical generator ${\cal L}$  which 
includes two distinct contributions: a coherent preserving term associated with an Hamiltonian operator, and a purely dissipative one, associated with a Lindblad super-operator term.
For assigned intensity of the latter our goal is to determine how $\tau_{ent}$ 
varies when increasing the intensity  of the former, in order to 
understand whether Hamiltonian corrections always help preserving entanglement, and more generally to unveil the interplay between purely dissipative and coherence preserving contributions in the Lindblad generator. 
Naively one would expect that a predominance
of the Hamiltonian term would tend to increase the survival time of the entanglement. However for the schemes we have considered this is not the case: the minimal value of $\tau_{ent}$ being
reached for a non zero value of the Hamiltonian intensity.  

In our analysis we shall formally identify $\tau_{ent}$ with the smallest time 
interval  after which  the dynamics associated with the selected ${\cal L}$ 
becomes an Entanglement-Breaking (EB) quantum 
channel~\cite{horodecki03,holevo99}. 
This choice makes sure that, irrespectively from the initial conditions, no entanglement between the system of interest and any possible ancillary system will survive after $\tau_{ent}$.
Conclusive results are presented for the case of qubit systems and for continuous variable systems evolving under the action of Gaussian noise. 

The presented material is organized as follows:
we start in Sec.~\ref{sec:tstar} introducing the formal definition of 
entanglement survival time for generic open quantum system dynamics and 
review some basic properties of dynamical semigroups. After presenting a detailed analysis of the general properties of the 
entanglement survival time in Sec.  
\ref{GENERAL}, we focus on some models. Notably, in 
Sec.~\ref{sec:examples} we present some results dealing with qubit systems while in Sec.~\ref{Sec:GAUS} we extend the analysis to the case of Gaussian Bosonic channels. 
Conclusions and final remarks are presented in Sec.~\ref{SEC:CON}, while technical derivations are presented in the Appendix. 

\section{Maximum Entanglement Survival Time}
\label{sec:tstar}

Consider a quantum system $A$ that is evolving under the noisy influence of an external environment $E$,
whose action we represent by means of a continuous, one-parameter family $\{\Phi_{t,0}\}_{t\geq 
0}$  of 
completely positive, trace-preserving (CPt) linear super-operators~\cite{kraus_book,holevo_book,HOLEREV}.
Assume next that  at  $t=0$, $A$ is initialized 
 into  a (possibly entangled) joint 
state $\rho_{AB}(0)$ with an ancillary system $B$ which, without loss of generality we assume to be isomorphic with $A$, and 
which does not couple with $E$. 
In this setting we define $t^*(\rho_{AB}(0))$ 
the minimum temporal evolution time $t$ at which
  no entanglement can be found in the associated  evolved density matrix 
 \begin{eqnarray} 
 \rho_{AB}(t) =( \Phi_{t,0} \otimes \mathrm{id}_{B}) [ \rho_{AB}(0)]\;, \label{EVO} 
 \end{eqnarray} 
 ($\mathrm{id}_{B}$ being the identity super-operator on $B$), i.e. 
the quantity
\begin{eqnarray} \label{DEFTprima} 
t^*(\rho_{AB}(0))  := \min \{ {t\geq 0} \; \mbox{s.t.} \; \rho_{AB}(t) \in  {\frak S}_{sep}(\Ham_{AB})
\},
\end{eqnarray} 
with  ${\frak S}_{sep}(\Ham_{AB})$  the subset of separable states of $AB$. 
As explicitly indicated by the notation  
the expression in (\ref{DEFTprima})  is a function of the chosen initial state $\rho_{AB}(0)$: it runs from the minimum value $0$ (attained when $\rho_{AB}(0)$ is an element of ${\frak S}_{sep}(\Ham_{AB})$) to a maximum value 
\begin{eqnarray} \label{DEFTseconda} 
\tau_{ent} := \max_{\rho_{AB}(0)\in  {\frak S}(\Ham_{AB})} t^*(\rho_{AB}(0)) \;,
\end{eqnarray} 
which only depends upon the properties of the maps $\{\Phi_{t,0}\}_{t\geq 
0}$ and which can be 
equivalently expressed as the smallest time  $t$ for which 
$\Phi_{t,0}$ becomes EB, i.e. 
\begin{eqnarray} \label{DEFTsecondaUNO} 
\tau_{ent} = \min \{ t\geq 0 \; \mbox{s.t.} \; \Phi_{t,0} \in EB \}\;.
\end{eqnarray} 
Since it defines the maximum time interval on which we are  guaranteed to have some entanglement between $A$ and $B$ under the evolution~(\ref{EVO}),  we shall refer to 
$\tau_{ent}$  as  the ``{entanglement 
survival time}" (EST) of the selected dynamical process. 
Notice however that 
if the maps $\{\Phi_{t,0} \}_{t\geq0}$ exhibit a strong non-Markovian character
inducing a significative back-flow of information into the system temporal evolution~\cite{BLP, BLC, Mazzola, bellomo, turku},
 nothing prevents the possibility  that 
entanglement between $A$ and $B$ will re-emerge at some time $t$ greater than $\tau_{ent}$. 
The same effect however cannot occur 
in the case of  Markovian or weakly non-Markovian models for which 
instead one has
\begin{eqnarray} \label{DD}
\Phi_{t,0} \in EB  \quad \mbox{for all $t\geq \tau_{ent}$,}\end{eqnarray}
meaning that the
$AB$ entanglement is lost forever at time~$\tau_{ent}$. 
Following the approach of Refs.~\cite{BREUER} these two special classes of 
processes are  characterized by families $\{\Phi_{t,0}\}_{t\geq0}$ whose 
elements fulfil the 
CP-divisibility or P-divisibility condition  respectively, i.e. 
\begin{eqnarray} \label{LAMBDA} 
 \Phi_{t,0} = \Lambda_{t,t'} \circ \Phi_{t',0}\;,  \quad \forall t\geq t'\geq 0\;, 
 \end{eqnarray} 
where ``$\circ$" indicates the composition of super-operators and where the connecting element $\Lambda_{t,t'}$ are CP (Markovian processes) or 
simply positive transformations (weakly non-Markovian processes). 
 Equation~(\ref{DD}) can then be derived by setting $t' =\tau_{ent}$ in (\ref{LAMBDA}) and exploiting the fact that  the composition of an EB channel with a CP, or just positive, map is still EB.
\\

An important subclass of 
 Markovian (CP-divisible) processes is provided by the so called dynamical semi-groups, 
 characterized by channels $\{\Phi_{t,0}\}_{t\geq 0}$ which are  invariant under translations of the time
coordinates or, equivalently, by connecting maps  which are time homogeneous,  
 i.e. 
 \begin{eqnarray} \label{FFD}
 \Lambda_{t,t'}=\Lambda_{t-t',0}=\Phi_{t-t',0}\;, \quad \forall t\geq t'.
 \end{eqnarray} 
Accordingly defining $\Phi_{t}:= \Phi_{t,0}$,  Eq.~(\ref{FFD}) allows us to recast~(\ref{LAMBDA}) in terms of the following 
semigroup identity 
\begin{equation}
\Phi_{t}\circ\Phi_{\Delta t} =\Phi_{\Delta t}\circ\Phi_{t}  = \Phi_{t+\Delta t}, \quad \forall t, \Delta t \geq 0, 
\label{eq:semigroup_prop}
\end{equation}
which ultimately yields to a first order differential equation
\begin{equation}
	\dot{\Phi}_t = \Lag \circ \Phi_t, \qquad \Phi_0=\rm id\;,
	\label{eq:markov}
\end{equation}
driven by a Gorini, Kossakowski, Sudarshan, Lindblad (GKSL) generator $\Lag$~\cite{gorini76,lindblad76}.
The latter
admits  a standard decomposition in terms of two competing terms:
a coherence  preserving contribution gauged by an Hamiltonian term governed by a self-adjoint operator $H$ and by a purely dissipative term ${\cal D}$ governing the irreversible process. In the specific, we have 
\begin{eqnarray}  \label{DECO} 
\Lag[\;\cdot\;]
 =  \gamma \; {\cal D}[\;\cdot\;] -i\omega\;  [H,\;\cdot\;]_- \;,
 \end{eqnarray} 
 with 
     \begin{equation}
{\cal D}[\;\cdot\;] = 
\sum_{j=1}^{d^2-1}\left(L_j[\;\cdot\;] L_j^\dagger -{1\over2} \left[ 
L_j^\dagger 
\;L_j, \;\cdot\;\right]_+
\right),
\label{eq:lindblad}
\end{equation}
 the sum  running over a set of no better specified (Lindblad)
operators $\{ L_j\}_{j}$, and 
 the symbols $[\;,\;]_{\pm}$ indicating the commutator ($-$) and 
anti-commutators ($+$)  brackets, respectively ($d$ being the dimension of $A$).
 In Eq.~(\ref{DECO}) the quantities  $\omega, \gamma\geq 0$ have  dimension of a frequency and gauge the time scale and the relative
  strengths of the two competing dynamical mechanisms that act on $A$: 
  accordingly we shall refer 
$\omega$ as the (unitary) {driving parameter} and to $\gamma$ as the {damping 
parameter} (herewith and in the following we set $\hbar=1$ for the sake of convenience).

As Eq.~(\ref{eq:markov}) admits a formal integration 
\begin{eqnarray} 
\Phi_t = e^{t \Lag}\;,  \label{MAPPEL} 
\end{eqnarray}   it is clear that the {EST} of a dynamical semigroup must be a functional 
of its generator, i.e. 
\begin{eqnarray} \tau_{ent} = \tau_{ent}({\cal L})\;. \label{FUN} 
\end{eqnarray}
Analyzing such dependence is the aim of the present work. More precisely, for fixed $H$ and ${\cal D}$ we are
interested in studying in which way the  parameters $\omega$ and $\gamma$ that measure the relative
``strengths"
of the Hamiltonian and the dissipative contributions 
of $\Lag$ affect the value of $\tau_{ent}$. 
Intuitively one would aspect that
larger incidence of the first mechanism with respect to the second one would yield longer values of the corresponding {EST}. Interestingly enough it turns out that this is not always the case: as we shall explicitly see, in some circumstances 
the presence of a non zero value of the Hamiltonian parameter $\omega$ induces a drastic reduction of the {EST} of the model.

\section{Evaluating {EST} for dynamical semigroup}  \label{SEC:IMPO} 
In this section we analyze a few examples of dynamical semigroups and compute their associated {EST}.
We start in Sec.~\ref{GENERAL} by presenting some general properties of the functional (\ref{FUN}). In Sec.~\ref{Sec:QUBIT} 
we focus instead on the special cases of qubit systems which allow for an almost complete analytical treatment. Finally in Sec.~\ref{Sec:GAUS}
we discuss the problem in the context of Gaussian Bosonic Channels.

\subsection{Preliminary observations}\label{GENERAL} 
In the study of  the functional (\ref{FUN})
 some structural properties of the 
GKSL generator should be taken into consideration.
First of all, an almost immediate consequence of our definitions is the following scaling law
\begin{eqnarray}\label{prop}
\tau_{ent}( q {\cal L})= \;  \tau_{ent}({\cal L})/q \end{eqnarray} 
that holds for all $q\geq 0$ and for all ${\cal L}$. Hence for fixed $H$ and ${\cal D}$ we can write 
\begin{eqnarray} \label{NEWTAU}
\tau_{ent}({\cal L})={\cal T}_{ent}(\kappa) /\gamma\;,  \end{eqnarray}  
where 
\begin{eqnarray} \kappa :=\omega/\gamma\;,\label{DEFKAPPA}
\end{eqnarray} 
 is the
ratio of the driven and damping constants of the model, and  
 ${\cal T}_{ent}(\kappa)$ is a dimensionless quantity associated with the 
(dimensionless) GKSL generator   ${\cal D}[\,\cdot\,] -i\kappa\;  
[H,\,\cdot\,]_-$.
Next we remind that  the decomposition (\ref{DECO}) is not unique as  $H$  and the associated  Lindblad operators $\{ L_j\}_j$ can be freely 
redefined according to the transformations 
\begin{eqnarray}
 H &\rightarrow&  H' =  H + \frac{{1}}{2i \kappa} \sum_j ( c_j^* L_j - c_j L_j^\dag) + b\;,  \nonumber  \\
  L_j &\rightarrow&  L_j' =  L_j + c_j\;,  \label{RESCALING} 
\end{eqnarray} 
with $c_j$ being complex numbers and $b$ being an arbitrary real parameter~\cite{breuer_petruccione_book}, and where
the ratio $\kappa$ on the first term accounts for the strength parameters 
$\gamma$ and $\omega$. 
While the term  $b$ plays no role in the derivation (it gets cancelled when entering the commutation
brackets), the coefficients $c_j$ induce a non trivial symmetry into the model  that we fix by
forcing the $L_j$ to be traceless.

A further symmetry of the problem arises from the fact that 
 local unitary transformations cannot create nor destroy 
entanglement~\cite{nielsen_chuang_book}. Accordingly 
the {EST} 
of an arbitrary (not necessarily Markovian) process $\{\Phi_{t,0}\}_{t\geq 0}$ is invariant under  transformations of the form 
\begin{equation}
 \Phi'_{t,0} = {\cal V}_t\circ\Phi_{t,0}\circ{\cal U}_t, \label{FFDE}
\end{equation} 
where ${\cal U}_t[\,\cdot\,] =U_t[\,\cdot\,] U_t^\dagger$ and ${\cal 
V}_t=V_t[\,\cdot\,] 
 V_t^\dagger$ represent  unitary conjugations induced by the (possibly 
time-dependent) operators $U_t$ and $V_t$, respectively.
At the level of dynamical semigroup this 
translates into the following identity \begin{equation}
\tau_{ent} (\Lag) = \tau_{ent}({\cal U}^{-1}\circ\Lag\circ{\cal U}), 
\label{eq:unitary_conj}
\end{equation}
that holds for 
a generic (time-independent) 
unitary conjugation ${\cal U}$.
Equation~(\ref{eq:unitary_conj}) can be easily verified by 
noticing that given the semigroup $\Phi_t$ generated by 
$\Lag$, and the semigroup $\Phi'_t$ generated by $\Lag'={\cal 
U}^{-1}\circ\Lag\circ{\cal U}$,  the two are connected 
as in (\ref{FFDE}) by setting ${\cal V}_t={\cal 
U}^{-1}$ and ${\cal U}_t={\cal 
U}$.
Notice also that invariance of the {EST} under (\ref{FFDE}) can be used to explicitly verify 
that in the evaluation of such parameter it does not matter whether we integrate~(\ref{eq:markov})
directly or by passing through the standard interaction picture.
Indeed by setting
 ${\cal U}_t=\mathrm{id}$ and  identifying ${\cal V}^{-1}_t$ with  the evolution
induced by the Hamiltonian $H$ of (\ref{DECO}),  the integration 
of~(\ref{eq:markov}) in the standard interaction picture 
  can be seen as  a special
instance of~(\ref{FFDE}), with $\Phi^{'}_{t,0}$  being the non-homogenous 
Markovian process characterized by 
the  time dependent generator 
 ${\cal L}'_t = {\cal V}_t \circ {\cal D} \circ {\cal V}^{-1}_t$.

\subsection{Qubit systems}\label{Sec:QUBIT}

In Ref.~\cite{horodecki03} it has been established that determining whether a 
given CPt map $\Phi$ is EB,  is equivalent 
 to check if its  associated  Choi-Jamio\l kowski 
state $\rho_{AB}^{(\Phi)}$~\cite{choi75,jamiolkowski72} is separable or not. For finite dimensional systems  the latter is defined as the 
output density matrix generated by $\Phi$ when acting locally on a maximally entangled state, i.e. 
\begin{eqnarray}
\rho_{AB}^{(\Phi)} &=& (\Phi \otimes \mathrm{id}_B) [\ket{\Omega}_{AB}\bra{\Omega}] \;, \label{CHOI} \\ 
|\Omega\rangle_{AB} &:=&\frac{1}{\sqrt{d}} \sum_{k=1}^d \ket{k}_A \otimes \ket{k}_{B}\;,  \label{MAX} 
\end{eqnarray}
where $d$ is the dimension of $A$, and where  for $Q=A,B$,  $\{|k\rangle_Q\}_{k=1,\cdots,d}$ is an orthonormal basis of the system $Q$. 
A direct consequence of this fact is that 
 the maximum in~(\ref{DEFTseconda}) is always attainable on the pure state~(\ref{MAX}), i.e. that 
 $\tau_{ent}({\cal L})$ of a semigroup $\{ \Phi_t\}_{t\geq 0}$ can be found as the minimum value of $t$ for which $\rho_{AB}^{(\Phi_t)}$ becomes separable. 
 If $A$ is a qubit, i.e. if $d=2$, we can address this task by exploiting  the positive partial transpose (PPT) criterion~\cite{horodecki96,peres96} 
 which states that $\rho_{AB}^{(\Phi_t)}$ is separable if and only if 
its partial transposes  (say $[\rho_{AB}^{(\Phi_t)}]^{T_B}$) is non negative, i.e. if and only if all its eigenvalues are greater than or equal to $0$. By 
continuity, $\tau_{ent}$ can then be also identified as the smallest $t$ which 
nullifies the determinant of $[\rho_{AB}^{(\Phi_t)}]^{T_B}$, i.e. 
\begin{equation}
\tau_{ent} = \min\{ t\geq 0, \; \mbox{s.t.} \;  \det([\rho_{AB}^{(\Phi_t)}]^{T_B}) = 0\}\;,
\label{eq:finite_dim}
\end{equation}
or, equivalently, as the smallest $t$  which nullifies the corresponding 
negativity of entanglement~\cite{vidal02}, i.e. 
\begin{equation}
\tau_{ent} = \min\{ t\geq 0, \; \mbox{s.t.} \;  \mathcal{N}(\rho_{AB}^{(\Phi_t)}) = 0\}\;,
\label{eq:finite_dim11}
\end{equation}
where given $\rho_{AB}$ a generic state we have 
\begin{equation} \label{NEGA}
\mathcal{N}(\rho_{AB}) = \frac{1}{2} \sum_{\ell} ( |\lambda_\ell| - \lambda_\ell )\;,
\end{equation}
with $\{\lambda_\ell\}_\ell$ being the eigenvalues of  $\rho_{AB}^{T_B}$. 
It is worth observing that  since the negativity of entanglement
is an entanglement monotone~\cite{vidal00} (the higher its values the higher is the entanglement present in the system), the 
 function $\mathcal{N}(\rho_{AB}^{(\Phi_t)})$ 
can also be used to monitor how the entanglement gets degraded before completely disappearing at $\tau_{ent}$. 
Furthermore we notice that 
 for $d>2$ where the PPT criterion   provides a sufficient but not necessary condition for separability, the terms on the right-hand-side of 
 Eqs.~(\ref{eq:finite_dim}) and (\ref{eq:finite_dim11}) provide an upper bound for $\tau_{ent}$.  
\\

\paragraph{Asymptotic analysis} 
While for $d=2$ determining the eigenvalues of $[\rho_{AB}^{(\Phi_t)}]^{T_B}$ is
always possible in principle,
extracting $\tau_{ent}$ from Eq.~(\ref{eq:finite_dim}) or~(\ref{eq:finite_dim11})  requires in general
to solve a transcendental equation. 
 As a result, only in a few cases it 
is possible to carry out the entire analysis analytically and one has to resort
to numerical methods, for instance to the Newton-Raphson method, which we shall 
employ extensively in the following sections (in particular in the plots shown in Figs.~\ref{fig:BHF_Tent}, ~\ref{fig:GAD} and \ref{fig:tstar_ampli}). Yet by inspecting the 
 asymptotic behaviour of the spectrum of $[\rho_{AB}^{(\Phi_t)}]^{T_B}$ (a relatively simple task) it can 
often be inferred whether the entanglement transmission time of a given dynamical semigroup  is finite or 
infinite. Consider in fact the case where the process admits  a single relaxation state, i.e. 
\begin{eqnarray} 
\lim_{t\rightarrow \infty} \Phi_t (\rho_A(0)) = \bar{\rho}_A \;, \qquad 
\forall \rho_A(0) \in {\frak S}(\Ham_{A})\;.
\end{eqnarray} 
with $\bar{\rho}_A$ being determined by the identity
 ${\cal L}[\bar{\rho}_A]=0$. 
Accordingly the  Choi-Jamio\l kowski will converge to the following separable state 
\begin{eqnarray} 
\label{RES1} 
 \rho_{AB}^{(\Phi_\infty)}  := \lim_{t\rightarrow \infty} \rho_{AB}^{(\Phi_t)} = \bar{\rho}_A \otimes \openone_B/2\;, 
\end{eqnarray} 
which implies 
\begin{eqnarray} 
\label{RES2} 
\det([\rho_{AB}^{(\Phi_\infty)}]^{T_B}) = \frac{\det(\bar{\rho}_A)}{4}
\geq 0\;. 
\end{eqnarray} 
Suppose hence that $\det(\bar{\rho}_A)>0$, which always happen unless the fixed point $\bar{\rho}_A$ is a pure state.
Then,  considering that for $t=0$ one has 
$\det([\rho_{AB}^{(\Phi_0)}]^{T_B})= \det(\ket{\Omega}_{AB}\bra{\Omega}^{T_B}) =-1/16$,  
 by a simple continuity argument it follows
 that 
 the function $\det([\rho_{AB}^{(\Phi_t)}]^{T_B})$ must cross $0$ at some finite time $t$ which, 
via (\ref{eq:finite_dim}),  corresponds to  the {EST} of the problem. 
If on the contrary we have $\det(\bar{\rho}_A)=0$, i.e. if $\bar{\rho}_A$ is pure, 
the continuity argument cannot be applied and the system may exhibit a divergent 
value of the {EST}, i.e. the associated dynamical semigroup becomes EB only asymptotically. 
Borrowing from the terminology introduced in Ref.~\cite{entdeath} we can hence conclude
that the purity of the relaxation state $\bar{\rho}_A$ provide a sufficient criterion for
determining whether the associated dynamical semigroup induces entanglement sudden death (ESD): 
\\

{\bf ESD Criterion:} {\it Dynamical semigroups admitting 
a non pure density matrix as relaxation state, are characterized by a finite value of the EST.} 
\\

We conclude by stressing  that while explicitly discussed for the qubit case scenario, it is clear that the above argument 
holds true for system $A$ of arbitrary dimension $d$, the only difference being 
associated with the fact that now the r.h.s. terms of Eq.~(\ref{RES1}) and (\ref{RES2}) get replaced
respectively by $\bar{\rho}_A \otimes \openone_B/d$ and  ${\det(\bar{\rho}_A)}/{d^d}$.
\\

\section{Models}\label{sec:examples}
 In this section we will explicitly explore the behavior of the $\tau_{ent}$ for different prototypical model for finite and infinite dimensional systems.
\subsection{Phase-flip qubit channels} 
As a first example of a dynamical semigroup acting on a qubit we consider
the case of GKSL generators $\Lag$~(\ref{DECO}) having and arbitrary  Hamiltonian term and  a unique
 Lindblad  operator $L$ which is Hermitian.
For a proper choice of the drift and the damping coefficients $\omega$, $\gamma$, and 
invoking the gauge freedom (\ref{RESCALING}) to make $L$ traceless, 
the most general example of such processes can be described 
by setting $L=Z/\sqrt{2}$ and taking $H = \hat{n}\cdot\vec{\sigma}$ with 
 $\hat{n} := 
(\sin\theta \cos\varphi , \sin\theta\sin\varphi , 
\cos\theta)$,
being real a unit vector, and with $\vec{\sigma}:=(X,Y,Z)$ being the vector of Pauli matrices.
Equation~(\ref{DECO}) hence becomes
\begin{equation}
{\cal L}[\;\cdot\;]  = \frac{\gamma}{2}\,(Z[\;\cdot\;] Z - 
\mathrm{id}[\;\cdot\;]  ) -i\omega\;  [\hat{n}\cdot\vec{\sigma},\;\cdot\;]_-\;, 
\label{eq:BFH}
\end{equation} 
which, in the computational basis associated with the  eigenvectors of $Z$,  can be interpreted as a phase-flip noise process~\cite{nielsen_chuang_book}
affecting the qubit $A$ while the latter evolves in the presence of a driving field in the $\hat{n}$ direction. 
Invoking the  equivalence~(\ref{eq:unitary_conj}) the analysis can be further simplified by observing that  a proper  unitary rotation
along the $z$ axis can be used to bring $\hat{n}$ into the $xz$ plane while keeping the dissipator component invariant. 
Accordingly, without loss of generality, in our analysis we shall set equal to zero the azimuthal angle $\varphi$, restricting the analysis to Hamiltonian driving of the form 
\begin{equation}\label{DEFN} 
 \hat{n} = 
(\sin\theta , 0 , 
\cos\theta)\;.
\end{equation}

As a preliminary step let us first consider the scenario where no coherent driving is acting on the system  ($\kappa=0$), so that 
${\cal L} =\gamma {\cal D}$. 
By explicit integration of the system dynamics (see Appendix~\ref{PAPPA1})
one can easily verify that in this case 
the  negativity of entanglement~(\ref{NEGA}) of  the associated   Choi-Jamio\l kowski  state~(\ref{CHOI}) is equal to 
\begin{equation}\label{ENNE}
{\cal N}(\rho_{AB}^{(\Phi_t)}) \Big|_{\kappa=0} =  e^{-\gamma t}/2\;,
\end{equation}
which shows that 
the entanglement in the system is degraded exponentially fast, even though it is 
never completely broken, yielding a divergent value for the associated {EST}, i.e. using (\ref{DEFKAPPA}) and (\ref{NEWTAU}),
  \begin{eqnarray}
 {\cal T}_{ent}(0) = \infty\;. \label{PRIMO}
  \end{eqnarray} 
 The same result holds also for arbitrary $\omega$ and $\hat{n}$ pointing into 
the $z$ axis, i.e. $\theta=0$. In this case in fact
 passing into the interaction picture representation 
  the   driving term can be eliminated without affecting the dissipator
 making the former completely irrelevant for the computation of the {EST} (see 
 comments at the end of Sec.~\ref{GENERAL}).

\begin{figure}[t]
\includegraphics[width=\columnwidth]{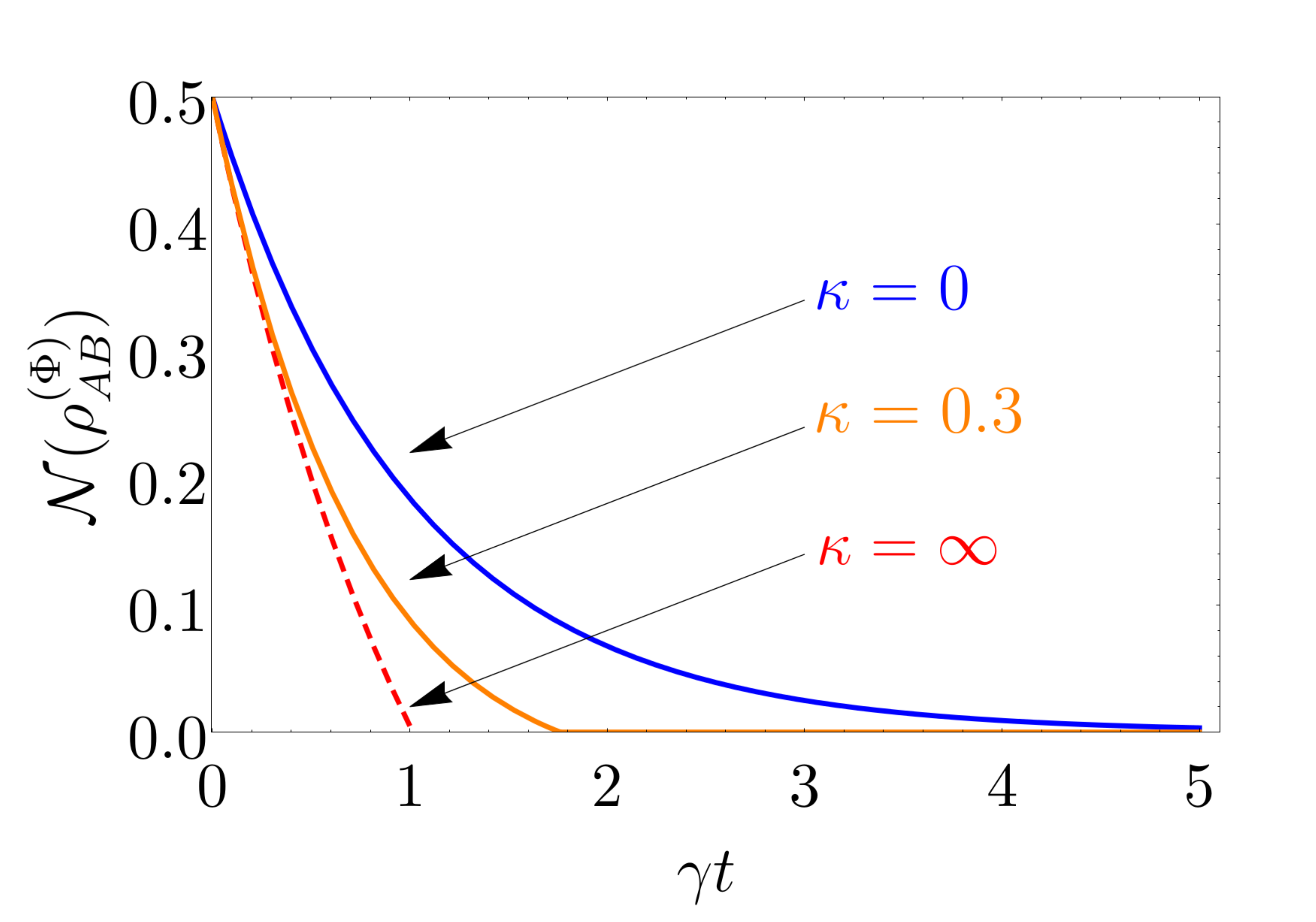}
\caption{(Color online) Temporal evolution of the negativity of 
entanglement (\ref{NEGA1}) of the phase-flip process \eqref{eq:BFH}
for different values of the ratio $\kappa=\omega/\gamma$ and for $\theta=\pi/2,\varphi=0$.}
\label{fig:BFZ_measures}
\end{figure}

\begin{figure*}[t!p]
\includegraphics[width=0.9\textwidth]{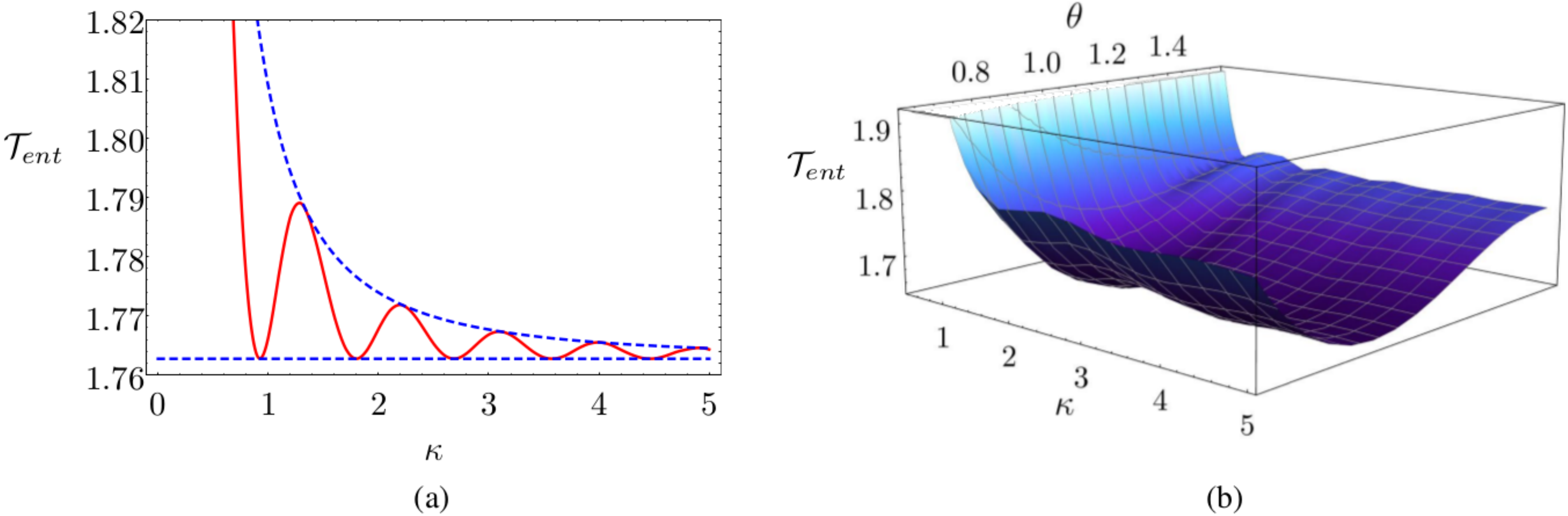}
\caption{ (Color online) Panel (a): plot of the rescaled {EST} functional  ${\cal T}_{ent}(\kappa)$ of
 the phase-flip channel~\eqref{eq:BFH} for $\theta=\pi/2,\varphi=0$ as a 
function of the driving/damping ratio $\kappa$  (red solid line), together 
with the bounds~\eqref{eq:tstar_bounds} (blue dashed lines). 
Panel (b):  3D plot of ${\cal T}_{ent}(\kappa)$ as a function
of $\kappa$ and of the rotation parameter $\theta$. Notice that 
again ${\cal T}_{ent}(\kappa)$ diverges for $\kappa\rightarrow 0$ and approaches a stationary value for $\kappa \rightarrow \infty$. The solid red line in Panel (a) and the 3D plot in Panel (b) have been generated numerically, exploiting the Newton-Raphson method.}
\label{fig:BHF_Tent}
\end{figure*}

The problem becomes more interesting  when we  take 
 $\hat{n}$ as a unit vector that  points  into the $x$ axis 
($\theta=\pi/2$), i.e. $H=X$. 
 Under these assumptions, 
 in the operator basis 
$\{E^{(00)},E^{(10)},E^{(01)},E^{(11)}\}$ formed by the external products 
$E^{(ij)}=\ket{i}\bra{j}$
of the computational basis, 
  the 
Lindbladian~\eqref{eq:BFH}  reads
\begin{equation}
\Lag = \gamma \begin{pmatrix}
0 & -i\kappa & i\kappa & 0 \\
-i\kappa & -1 & 0 & i\kappa \\
i\kappa & 0 & -1& -i\kappa \\
0 & i\kappa & -i\kappa & 0
\end{pmatrix}.
\label{eq:lag_BFZ}
\end{equation} 
By direct evaluation one can verify that for all $\kappa >0$ 
it admits as unique zero eigenvector the 
completely mixed state $\bar{\rho}_A = \openone_A/2$. Hence from the 
results of the previous section we can conclude that in these cases, at variance with the 
$\kappa=0$ scenario~(\ref{PRIMO}), the corresponding {EST} 
 must be finite yielding ESD~\cite{entdeath}. 
This is a rather remarkable fact
as it implies that by adding a unitary (coherent preserving) contribution to 
the dissipative dynamics induced by the phase-flip noise generator,  we can end 
up with a ``noisier'' evolution which becomes 
EB at a finite time. 
A more quantitative statement can be obtained by studying 
the 
negativity of entanglement, i.e. 
\begin{equation} \label{NEGA1} 
{\cal N}(\rho_{AB}^{(\Phi_t)}) = {e^{-\gamma 
t/2}\over2}\max\{Q_\kappa (\gamma t/2) -
\sinh(\gamma t/2) 
,0\},
\end{equation}
with 
\begin{equation}
Q_\kappa (\tau) := 
\sqrt{{\cosh^2({\tau}\sqrt{1-16\kappa^2})-16\kappa^2\over 
1-16\kappa^2}}. \label{QK} 
\end{equation}
The functional   dependence  of this quantity  upon the parameter  $\kappa$ 
is rather involved, still, as evident from the 
 plots  presented in Fig.~\ref{fig:BFZ_measures} it clearly emerges that the  entanglement present in the model tends to degrade faster as the driven/damping ratio increases.
According to Eq.~(\ref{eq:finite_dim11}) the associated {EST} can be determined by  identifying the zero's of (\ref{NEGA1}), i.e. 
solving   the transcendental  equation 
\begin{eqnarray} Q_\kappa (\gamma t/2) = \label{eq:maria} 
\sinh(\gamma t/2)\;, \end{eqnarray} 
which admits closed analytical solution for the two extremal cases  $\kappa=0$ and 
$\kappa\rightarrow\infty$. In particular for $\kappa=0$,  since 
$Q_0(\tau)=\cosh({\tau})$ Eq.~(\ref{eq:maria}) allows us to  recover the results 
anticipated in 
Eqs.~(\ref{ENNE}) and (\ref{PRIMO}). For  $\kappa=\infty$ instead
one has that  $Q_\kappa(\tau)$ converges to 
1,  allowing us to replace
 Eq.~(\ref{NEGA1}) with
\begin{equation}
 {\cal N}(\rho_{AB}^{(\Phi_t)}) \Big|_{\kappa=\infty}  =
{e^{-\gamma t/2}\over2} \max\{1-\sinh(\gamma t/2),0\},
\end{equation}
and yielding the following value for the  associated  rescaled {EST} functional (\ref{NEWTAU}) 
\begin{eqnarray} 
  {\cal T}_{ent}(\infty) ={\rm arcosh}(3)\;. 
\end{eqnarray} 
For the remaining choices of the driving/damping ratio $\kappa$  an approximate treatment 
 of (\ref{eq:maria}) allows us to write 
\begin{equation}
{\cal T}_{ent}(\kappa)  \simeq \begin{cases}
               W(1/4\kappa^2) & \kappa \simeq0,  \\ 
               2.5 -3.7\left({\kappa}-{1\over4}\right) 
+ 10.6\left({\kappa}-{1\over4}\right)^2 & \kappa \simeq \tfrac{1}{4}, \\ \\
               {\cal T}_{ent}(\infty)  + 
{1-\cos({\cal T}_{ent}(\infty) \sqrt{16\kappa^2-1})\over 
2\sqrt{2}(16\kappa^2-1)} & \kappa \simeq\infty,\\
              \end{cases} \label{CASES} 
\end{equation}
where $W$ is the Lambert function~\cite{coreless96} -- see 
Appendix~\ref{sec:perturb_th} for details.
Furthermore in the high driving regime  $\kappa \geq 1/4$ the following inequality can be established 
\begin{equation}
{\cal T}_{ent}(\infty)  \leq  {\cal T}_{ent}(\kappa)  \leq\mathrm{arcosh}\left(2-{
	1+16\kappa^2\over1-16\kappa^2}
\right).
\label{eq:tstar_bounds}
\end{equation}
In Fig.~\ref{fig:BHF_Tent}(a) we report a numerical solution of
Eq.~\eqref{eq:maria}, together with the bounds~\eqref{eq:tstar_bounds} which 
confirms the general tendency of the model in translating high level of unitary driving 
into a stronger entanglement suppression. 
A similar behaviour is  observed for intermediate values of $\theta$ in the
interval $[0,\pi/2]$ until it eventually diverges everywhere when $\theta$ 
approaches $0$: 
a numerical evaluation of the associated value ${\cal T}_{ent}(\kappa)$ is reported in 
Fig.~\ref{fig:BHF_Tent}.
\\ \\

\subsection{Generalized Amplitude Damping 
Process} 
As our next example  we focus on the case where
the dissipator ${\cal D}$ describes a generalized amplitude damping process (see e.g. Ref~\cite{depasquale12})
inducing  bosonic thermalization  effects on the qubit dynamics. 
It can be expressed as in Eqs.~(\ref{DECO}), (\ref{eq:lindblad}) 
by setting 
\begin{eqnarray}\label{GAD_gen}
L_1= \sqrt{N+1}\;  \sigma_- \;, \qquad L_2=  \sqrt{N} \; \sigma_+ \;,
\end{eqnarray} 
with $N$ being a non negative number that gauges the mean thermal photon number of the system environment and with $\sigma_{\pm}={1\over2}(X\pm iY)$ being ladder operators. 

In the absence of the driving term (i.e. $\omega=0$ or equivalently $\kappa=0$) the model can be easily 
integrated the generator taking the matrix form 
\begin{equation}
\Lag = \begin{pmatrix}
-\gamma_2 & 0 & 0 & \gamma_1 \\
0 & -{1\over2}(\gamma_1+\gamma_2) & 0 & 0 \\
0 & 0 & -{1\over2}(\gamma_1+\gamma_2) & 0 \\
\gamma_2 & 0 & 0 & -\gamma_1 \\
\end{pmatrix},
\label{eq:GAD_matrix}
\end{equation}
where for ease of notation $\gamma_1$ and $\gamma_2$ stands for
$\gamma_1 = \gamma (N+1)$ and $\gamma_2= \gamma N$. 
In this limit the process admits the density matrix 
\begin{equation}
\bar{\rho}_A 
 = 
{1\over 2N+1}\begin{pmatrix}
N+1 & 0 \\
0 & N
\end{pmatrix},
\end{equation}
as unique stationary solution, which for $N > 0$  is always not pure. 
For this choice of the parameter we can hence invoke the ESD criterion 
 to establish that 
 the model  must  exhibit a finite value of the
EST parameter.  
The negativity of entanglement can be computed as well leading to 
\begin{eqnarray}\label{NENGA} 
 {\cal N}(\rho_{AB}^{(\Phi_t)}) &=& {e^{-(2N+1)\gamma t/2}\over2}
 \\
&& \times  
\max\{A_N(\gamma t)-\sinh( 
(2N+1) \gamma t/2),0\}, \nonumber 
\end{eqnarray}
where we have introduced the function
\begin{equation}
A_N(\tau)=\sqrt{{1\over2} + 
{
4N(N+1)+ \cosh((2N+1)\tau)\over2(2N+1)^2}}.
\label{eq:amplitude}
\end{equation}
For $N=0$ (purely lossy dynamics) the above expression reduces to ${\cal N}(\rho_{AB}^{(\Phi_t)}) =\,e^{-\gamma t}/2$ and the process never reaches the EB regime yielding a divergent value of $\tau_{ent}$, i.e.
\begin{equation}
 {\cal T}_{ent}(0)\Big|_{N=0}   =\infty \;. 
	\label{eq:tstar_gad_dissss1111}
\end{equation}
For $N> 0$  instead, determining the zero of the r.h.s. term of Eq.~(\ref{NENGA})
shows that 
 the EST is finite and expressed as in (\ref{NEWTAU}) with 
\begin{equation}
 {\cal T}_{ent}(0)  =\frac{1}{2N+1}  {\rm 
	arcosh}\left(1+{(2N+1)^2\over2N(N+1)}\right).
	\label{eq:tstar_gad_diss}
\end{equation}

\begin{figure*}[t!p]
\includegraphics[width=.9\textwidth]{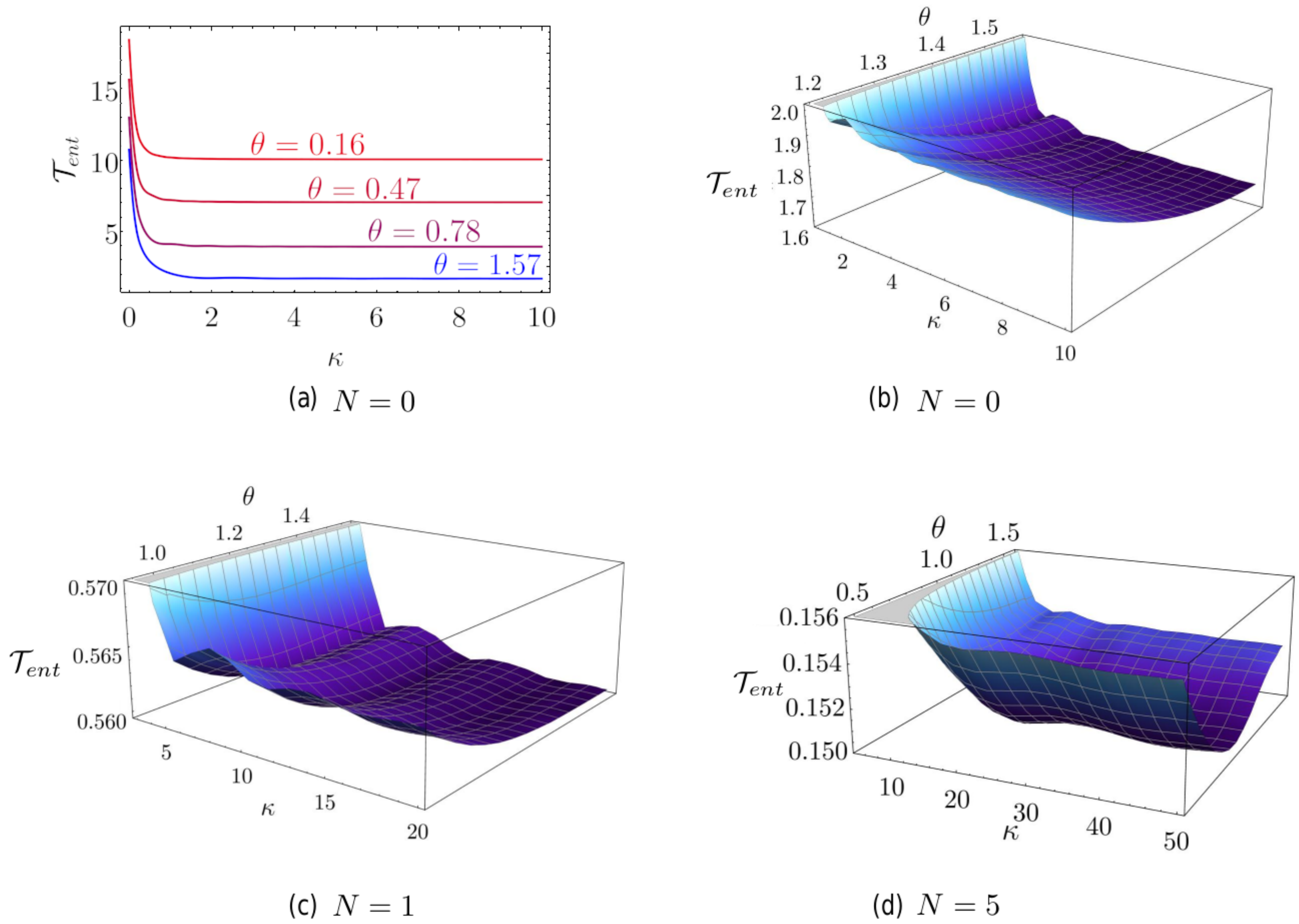}
\caption{(Color online) Panel (a): Plot of the rescaled EST  as a function of 
ratio $\kappa$ for  the amplitude damping process ($N=0$),  for several values 
of the parameter $\theta$ that determines the orientation of the driving 
Hamiltonian term. Panels (b), (c) and (d): Rescaled EST as a function of 
$\kappa$ and of the rotation parameter $\theta$ associated to the generalized 
amplitude damping process, for different values of the environment mean of 
photons number $N$. The plots shown in all the panels have been generated numerically employing the Newton-Raphson method.}
\label{fig:GAD}
\end{figure*}

Let us now allow for a non-zero ($\omega>0$) driving term $H = 
\hat{n}\cdot\vec{\sigma}$. In analogy with the phase-flip process, if we set 
$\hat{n}=(0,0,1)$ the Hamiltonian part of $\Lag$ can be eliminated by passing into the
interaction picture representation, therefore the EST does not depend on $\omega$. Also, exploiting the unitary 
invariance~(\ref{eq:unitary_conj}), the azimuthal angle 
$\varphi$ can be set to 0 without loss of generality, leaving us only with the 
dependence on $\theta$ to be resolved.
In Fig.~\ref{fig:GAD} we report the entanglement transmission curve for 
different values of the rotation parameter $\theta$ and the mean number of 
photons $N$.
We notice how once more the entanglement transmission time decreases with the driving/damping ratio $\kappa$. The 
qualitative behaviour of the curves is similar to those observed for the phase-flip model. In particular we notice that at fixed
$\kappa$, the values of ${\cal T}_{ent}(\kappa)$ develop a nontrivial minimum for intermediate values of $\theta \in ]0, \pi/2[$,
the effect being more evident at large $N$. 
\\ 

\subsection{The Depolarizing Process}
The last example we consider is the
 depolarizing process generated by a GKSL generator with the following three
Lindblad operators
\begin{eqnarray} 
L_1= X/2\;, \quad L_2= Y/2\;, \quad L_3= Z/2\;, 
\end{eqnarray} 
leading to a dissipator of the form 
\begin{equation} \label{DISSS} 
{\cal D}[\;\cdot\;]  =  {1\over4}(X[\;\cdot\;]
X+Y[\;\cdot\;] Y+ Z[\;\cdot\;] Z-3 \; \mathrm{id}[\;\cdot\;]).
\end{equation} 
In this case due to the highly symmetric structure of (\ref{DISSS})  any Hamiltonian 
contribution  can be eliminated by passing into the interaction picture without modifying the dissipator. 
More precisely, one can show that any purely Hamiltonian 
superoperator (i.e. one with $\gamma=0$) commutes with (\ref{DISSS}). This can be exploited so as to get rid of any functional dependence of EST 
 on $\omega$, and ultimately on $k$ as prescribed by \eqref{prop}:
\begin{equation}
 {\cal T}_{ent}(\kappa)   = {\cal T}_{ent}(0)\;,
	\label{eq:tstar_gad_new}
\end{equation}
for all $\kappa$. 
Neglecting hence $H$,  in the basis of 
the elementary matrices we observe that the generator becomes 
\begin{equation}
\Lag = {\gamma \over2}\begin{pmatrix}
-1 & 0 & 0 & 1 \\
0 & -2 & 0 & 0 \\
0 & 0 & -2 & 0 \\
1 & 0 & 0 & -1
\end{pmatrix},
\label{eq:ampdamp_gen}
\end{equation}
which, by direct exponentiation, leads to 
\begin{equation}
\Phi_t = {1\over2}\begin{pmatrix}
1+e^{-\gamma t} & 0 & 0 & 1-e^{-\gamma 
t} \\
0 & 2e^{-\gamma t} & 0 & 0 \\
0 & 0 & 2e^{-\gamma t} & 0 \\
1-e^{-\gamma t} & 0 & 0 & 1+e^{-\gamma t} 
\end{pmatrix}.
\end{equation}
Therefore, via a proper rearrangement of  the above matrix elements (divided by $2$), the Choi-Jamio\l kowski  reads
\begin{equation}
\rho_{AB}^{(\Phi_t)}= {1\over4} \begin{pmatrix}
1+e^{-\gamma t} & 0 & 0 & 2e^{-\gamma t} \\
0 & 1-e^{-\gamma t} & 0 & 0 \\
0 & 0 & 1-e^{-\gamma t} & 0 \\
2e^{-\gamma t} & 0 & 0 & 1+e^{-\gamma t}
\end{pmatrix}.
\end{equation}
The negativity of entanglement can then be computed as
\begin{equation}
 {\cal N}(\rho_{AB}^{(\Phi_t)})  = {e^{-\gamma t/2}\over2}\max\{e^{-\gamma 
t/2}-\sinh(\gamma t/2),0\},
\end{equation}
showing that the entanglement of the system is degraded, again, exponentially fast
with rescaled EST value given by the natural logarithm
\begin{equation}
 {\cal T}_{ent}(0) = \ln 3.
\end{equation}

\subsection{Gaussian Bosonic Channels}\label{Sec:GAUS} 
In this Section we address the case  of dynamical semigroups acting on
 infinite dimensional systems (continuous variables regime). In particular we 
shall focus on the special class of CPt maps which belongs to the set of 
Gaussian Bosonic channels~\cite{holevo_book,holevo01,gauss}, that we briefly review in Appendix~\ref{Sec:GAUSAPPENDIX}. Specifically, 
we consider the continous variables analog of the generalized amplitude damping process introduced earlier.
This process is described by 
 a GKSL generator~(\ref{DECO}) with  two Lindblad operators
\begin{eqnarray}  \label{LINDG} 
L_1= \sqrt{N+1}\;  a \;, \qquad L_2=  \sqrt{N} \; a^\dag \;, 
\end{eqnarray} 
with $N\geq 0$ representing the mean photon number of the environment and
$a$ and $a^\dag$ being, respectively, the annihilation and creation bosonic operators, fulfilling the canonical
commutation rule $[a,a^\dag]=1$. 
For the Hamiltonian part we take instead the most general quadratic operator which, without loss of generality, we parametrise as  
\begin{eqnarray} \label{HAMG} 
H = i \frac{(a^{\dagger})^2-a^2}{2}\sin\theta+a^\dagger 
a \; \cos\theta \;, 
\end{eqnarray} 
with  $\theta$ measuring the relative intensity of the squeezing term.

Consider first the case where no driving contribution is present (i.e. $\omega =0$). 
By explicit integration the associated CPt transformation $\Phi_t$ induced by ${\cal L}$ corresponds to a (single mode) Gaussian Bosonic channel 
which in the formalism detailed in Appendix~\ref{Sec:GAUSAPPENDIX} 
is described by the $2 \times 2$ real matrices 
\begin{equation}
	F_t=e^{-\gamma t/2}\,I, \quad 
G_t=(2N+1)(1-e^{-\gamma t})\,I\;. 
\end{equation}
This process belongs in the $C$ class and we can determine {its associated EST by finding solutions to the following equation 
\begin{equation} \label{IMPO111} 
 \det\left(G_t-{i\over2}(J+F^T_tJF_t)\right) = 0\;, 
\end{equation}
see Eq.~(\ref{IMPO1}). }
By explicit computation this yields the following value for the rescaled functional of~(\ref{NEWTAU}), i.e. 
\begin{equation} \label{TAUTAUE} 
  {\cal T}_{ent}(\kappa=0)   = \ln{4N+3\over4N+1},
\end{equation}
which, while being decreasing with $N$ as  its qubit counterpart~(\ref{eq:tstar_gad_diss}), at variance with the latter does not diverge when $N$ approaches zero, see Eq.~(\ref{eq:tstar_gad_dissss1111}).

\begin{figure*}[t!p]
\includegraphics[width=0.9\textwidth]{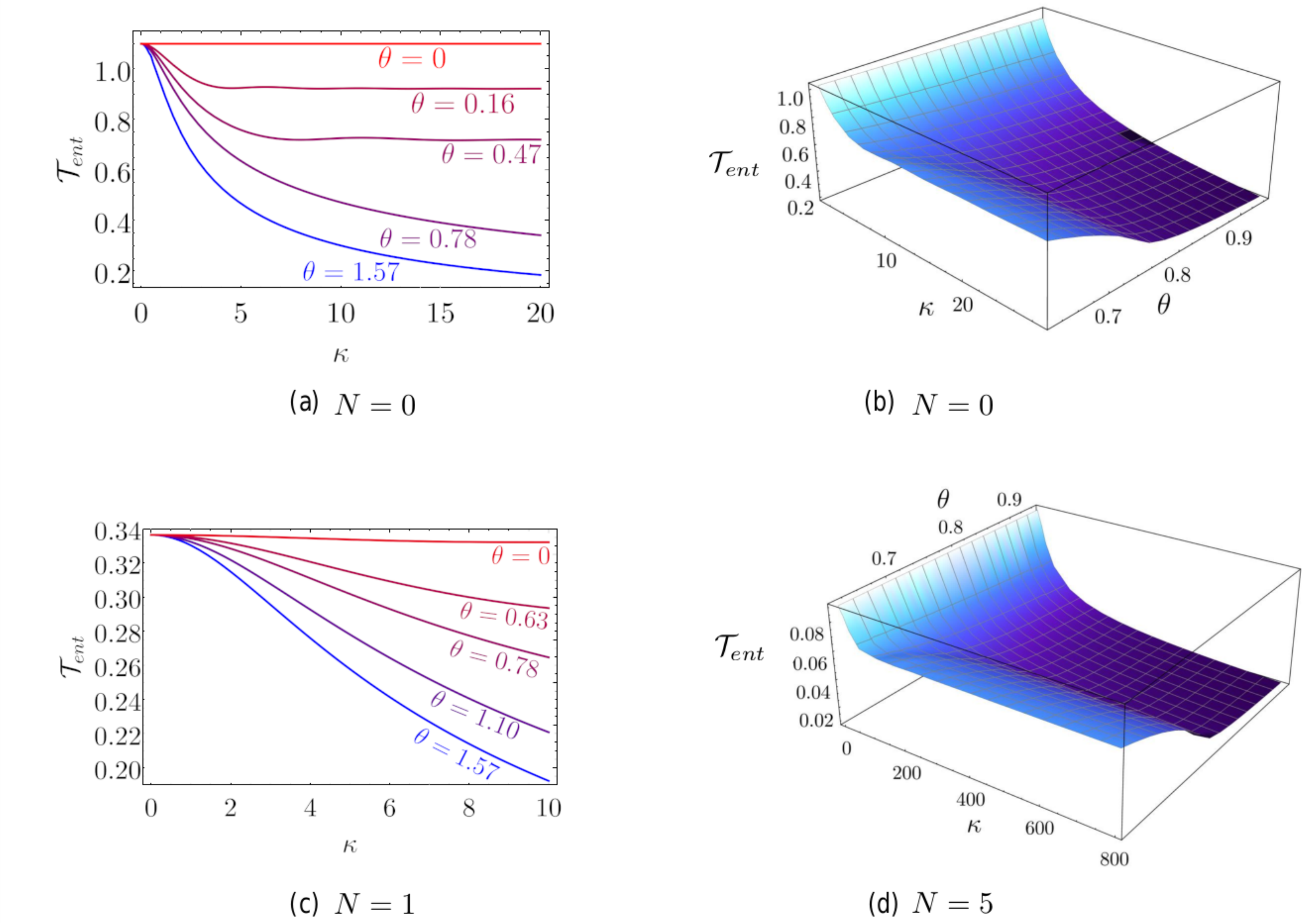}
\caption{\label{fig:tstar_ampli}(Color online) Panels (a) and (c): 
Rescaled EST ${\cal 
T}_{ent}$ as a function of the relative strength  $\kappa$ 
of dynamics associated to the Gaussian Bosonic Channel model defined by Eqs.~(\ref{LINDG}) and (\ref{HAMG}),
 for 
different values of the parameter $\theta$. Panels (b) and (d): 3D plot of the rescaled EST as a function of $\kappa$ 
and of the rotation parameter $\theta$. In all plots the values of ${\cal 
T}_{ent}$  have been obtained by numerically solving  Eq.~(\ref{QUESTAQUI}) of 
Appendix~\ref{APPFIN}, with the Newton-Raphson method.}
\end{figure*}

{Consider next the case of a non zero driving/damping ration, $\kappa>0$. For general $\theta$, Eq.~(\ref{IMPO111}) yields for the EST an 
equation analogous to~\eqref{eq:maria} which we report in Eq.~(\ref{QUESTAQUI}) of the Appendix
and whose numerical solution is exhibited  in Fig.~\ref{fig:tstar_ampli}.}

\begin{figure}[t!p]
	\includegraphics[width=.48\textwidth]{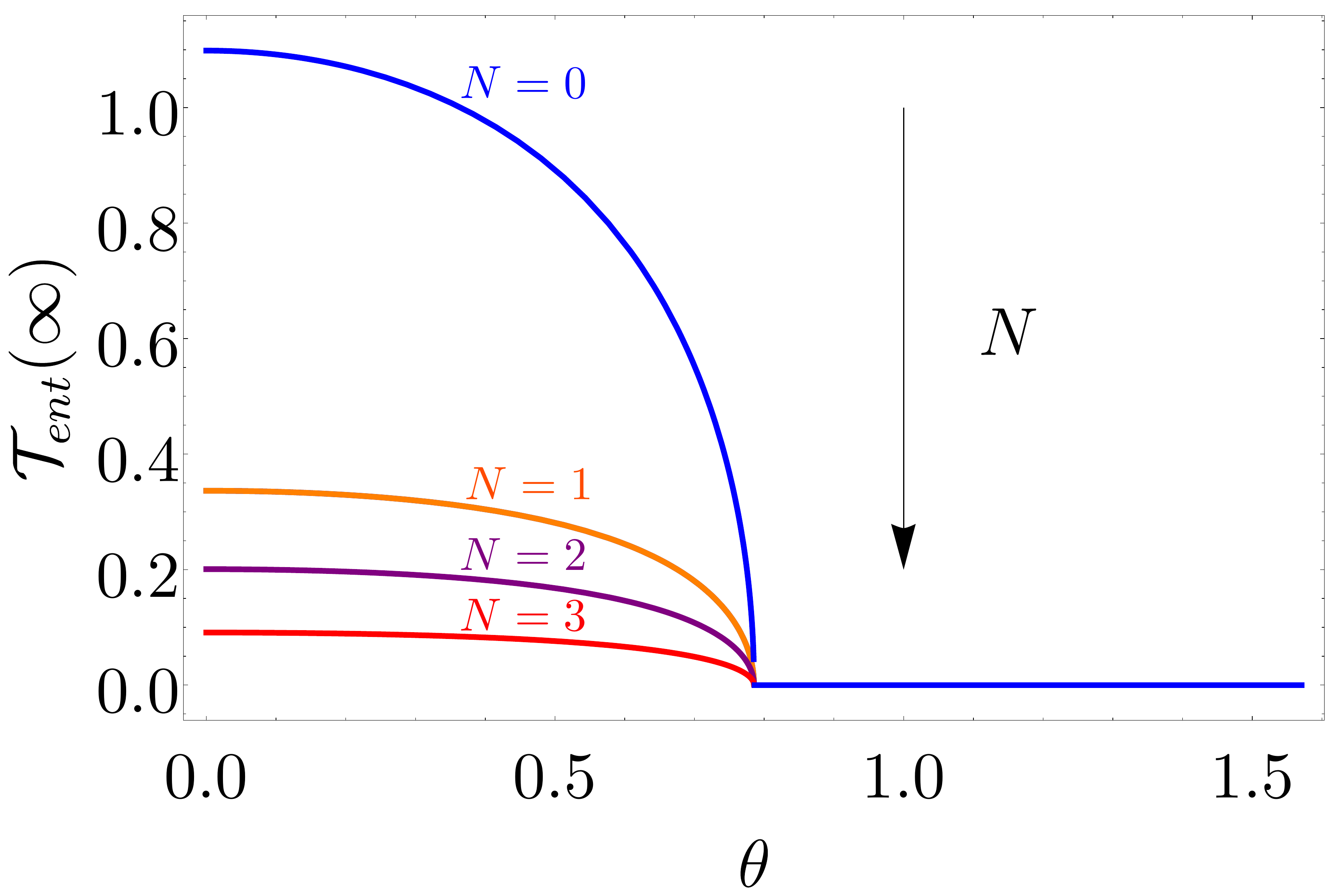}
	\caption{(Color online) Asymptotic form of the 
entanglement 
survival time~\eqref{gaussian_asymptote} of the Gaussian Bosonic 
Channel model defined by Eqs.~(\ref{LINDG}) and (\ref{HAMG})
as a function of $\theta$ for $\kappa\rightarrow\infty$.}\label{fig:lim_tstar}
\end{figure}

For $\kappa \simeq 0$ an approximate solution 
can be obtained in the following form 
\begin{align}
& {\cal T}_{ent}(\kappa)     \simeq 
\ln{4N+3\over4N+1} \nonumber \\ 
&+\kappa^2(2N+1)(1-\cos2\theta)\left[{4\over(4N+3)(4N+1)}-{\cal T}^2_{ent}(0)\right]\,.
\label{eq:gaussian_sqz}
\end{align} 
For large values of driving/damping ratio $\kappa$ instead, Eq.~(\ref{QUESTAQUI})   presents a critical behavior in $\theta$ (see  Fig.~\ref{fig:lim_tstar}).
{In particular for $\theta\in[0,\pi/4)$}
 the form of ${\cal T}_{ent}$ is similar to the finite dimensional case, exhibiting a {drop-oscillate-stabilize} pattern
which can be approximated by the function 
\begin{eqnarray} \label{FFFDD} 
	&&{\cal T}_{ent}(\kappa)   \simeq  {\cal T}_{ent}(\infty) 
	 \\  \nonumber
	 && \quad + {a\cos[2\kappa \sqrt{-\cos2\theta}\;  {\cal T}_{ent}(\infty )] + b- \cosh({\cal T}_{ent}(\infty))\over\sinh({\cal T}_{ent}(\infty))}\;,
\end{eqnarray}
 with ${\cal T}_{ent}(\infty)$ being the asymptotic value  defined as 
\begin{equation}\label{gaussian_asymptote}
 {\cal T}_{ent}(\infty)    
= {\rm arcosh}\frac{2(2N+1)^2+(8N(N+1)+3)\cos2\theta}
{2(2N+1)^2+(8N(N+1)+1)\cos2\theta},
\end{equation}
vanishing for $N\rightarrow+\infty$.
{For
 $\theta\in(\pi/4,\pi/2]$}, where eq.~\eqref{FFFDD} can have a 
non-zero imaginary part, instead the EST is
monotonically decreasing with $\kappa$, asymptotically vanishing in the large 
$\kappa$ regime. Since for
 $\theta\in(\pi/4,\pi/2]$ eq.~\eqref{QUESTAQUI} changes form, in this interval
the functional dependence is approximated by the 
function 
{
\begin{equation}
 {\cal T}_{ent}(\kappa)  \simeq {1\over2\kappa\sqrt{|\cos2\theta|}}{\rm 
	arcosh}\,{2\; \kappa^2\; \alpha(N,\theta)\;  \cos2\theta 
	\over 
	(2N+1)^2(1-\cos2\theta)}\;, \label{ENEW} 
\end{equation}
where the dimensionful quantity $\alpha$ is given by
\begin{align}
	\alpha(N,\theta) &= 
	2\gamma^2(2N+1)^2 + \gamma^2[8N (N+1) +3] \cos 2\theta \;.
\end{align}}

\section{Conclusions}\label{SEC:CON} 
The present paper focuses on the study of the
entanglement transmission time, defined as the time at which a dynamical 
process induced by the interaction with an external environment 
 becomes entanglement-breaking.
For the special case of time-homogeneous, Markovian systems, we analyze how this 
quantity is affected by the interplay between the dissipative and the driving 
contributions of the GKSL generator of the model. 
We provided both analytical and numerical results for some relevant examples of 
qubit evolution, described by the bit-flip and the amplitude damping channels.
In the simplest cases we evaluate also the negativity of entanglement, which 
quantifies the entanglement content of the semigroup output state, and therefore 
provide information also on the rate at which entanglement is being corrupted.
We noticed that the dependency of the entanglement transmission time from the 
damping and driving parameters reflects the form of the eigenvalues of  $\Lag$, the GKSL
generator of the quantum dynamical semigroup. The precise form of such 
dependency can be very complicated even in the simple cases considered, but 
generally it has been found that oscillations can appear in the entanglement 
transmission time. This 
happens in the finite-dimensional case when the eigenvalues of the generator 
acquire an imaginary part. In the infinite-dimensional case, we observe 
an oscillatory behaviour only for certain values of the rotation parameter.

Somewhat contrary to common intuition, our results clearly show that 
increasing the driving parameter, by tuning the weight of the unitary dynamics, does not always 
provide an advantage in the transmission of entanglement. Indeed, in the study 
cases considered it appears to be detrimental, making the transmission time 
drop, with the exception of a special driving direction, which makes the 
driving ineffective.
An intuitive explanation of this effect can be attempted by saying that the unitary rotations induced by 
the presence of  coherent preserving contributions in the GKSL generator, could effectively increase the detrimental effects  of the
dissipative ones, by broadening the range of their action in the phase space of the system.
In other words  by exposing the Hilbert space of the latter 
to attacks that can affect any possible subspaces, these rotations
boost the noise level inducing a 
``playing both sides of the fence"-effect where the system has no hidden paces where to
store the coherence it needs to maintain the entanglement with an eternal ancilla. 

\acknowledgements
We thank P. Mataloni e and F. Sciarrino for useful discussions.  
\appendix
\section{Formal integration of the Bit-Flip channel model} \label{PAPPA1}
 Setting  $\omega=0$ in the operator basis 
$\{E^{(00)},E^{(10)},E^{(01)},E^{(11)}\}$ formed by the external products 
$E^{(ij)}=\ket{i}\bra{j}$
of the computational basis, 
the Lindblad super-operator of the Phase-Flip channel model,
 takes the matrix form
\begin{equation}
\mathcal{\Lag} = \gamma \begin{pmatrix}
0 & 0 & 0 & 0 \\
0 & -1 & 0 & 0 \\
0 & 0 & -1 & 0 \\
0 & 0 & 0 & 0
\end{pmatrix},
\label{eq:bflip_gen}
\end{equation}
which gives   
\begin{equation}
\Phi_t = \begin{pmatrix}
1 & 0 & 0 & 0 \\
0 & e^{-\gamma t} & 0 & 0 \\
0 & 0 & e^{-\gamma t} & 0 \\
0& 0 & 0 & 1
\end{pmatrix},
\end{equation}
as the associated semigroup maps (\ref{MAPPEL}).
Adopting hence as the maximally entangled state (\ref{MAX}) the one constructed on the computational basis, 
i.e.  $|\Omega\rangle_{AB}\langle\Omega| = \sum_{j,j'=0,1} E_A^{(jj)} \otimes  E_B^{(j'j')}/4$, which for $d=2$ has the following matrix form
\begin{equation}
	\ket{\Omega}_{AB}\bra{\Omega} = {1\over2}\begin{pmatrix}
	1 & 0 & 0 & 1 \\
	0 & 0 & 0 & 0 \\
	0 & 0 & 0 & 0 \\
	1 & 0 & 0 & 1
	\end{pmatrix},
\end{equation}
the partial transpose of the corresponding  Choi-Jamio\l kowski state can likewise be expressed as
\begin{equation}
 [\rho_{AB}^{(\Phi_t)}]^{T_B}  = {1\over2}
 \begin{pmatrix}
1 & 0 & 0 & 0 \\
0 &0&  e^{-\gamma t} & 0 \\
0 & e^{-\gamma t} &0&  0 \\
0& 0 & 0 & 1
\end{pmatrix},
\end{equation}
having eigenvalues $1/2$ (twice degenerate) and 
$\pm e^{-\gamma t}/2$ which leads to (\ref{ENNE})
when replaced into (\ref{NEGA}). 

\section{Entanglement Negativity}
\label{sec:nega}
In this Section we provide some details for the derivation of negativity in the 
models described by the generator \eqref{eq:BFH} with $\hat{n}=(1,0,0)$ and the 
generator \eqref{GAD_gen} with $\omega=0$.

In the basis $\{E^{(00)},E^{(10)},E^{(01)},E^{(11)}\}$ the Lindbladian 
\eqref{eq:BFH} is represented by the matrix \eqref{eq:lag_BFZ}. By means of \eqref{eq:unitary_conj}, we can transform \eqref{eq:BFH} into the equivalent generator
\begin{equation}
\Lag = \gamma\begin{pmatrix}
-{1\over2} & 0 & 0 & {1\over2} \\
0 & -{1\over2} + 2i\kappa & {1\over2} & 0 \\
0 & {1\over2} & -{1\over2}-2i\kappa & 0 \\
{1\over2} & 0 & 0 & -{1\over2}
\end{pmatrix},
\end{equation}
which makes our analysis simpler. By taking the exponential, we have
\begin{widetext}
	\begin{equation}
	\Phi_t = e^{-{1\over2}\gamma t}\begin{pmatrix}
	\cosh{1\over2}\gamma t & 0 & 0 & \sinh{1\over2}\gamma t \\
	0 & \cosh{1\over2}\gamma t\sqrt{1-16\kappa^2}+{4i\kappa\sinh{1\over2}\gamma t\sqrt{1-16\kappa^2}\over\sqrt{1-16\kappa^2}} & {\sinh{1\over2}\gamma t\sqrt{1-16\kappa^2}\over\sqrt{1-16\kappa^2}} & 0 \\
	0 & {\sinh{1\over2}\gamma t\sqrt{1-16\kappa^2}\over\sqrt{1-16\kappa^2}} & \cosh{1\over2}\gamma t\sqrt{1-16\kappa^2}-{4i\kappa\sinh{1\over2}\gamma t\sqrt{1-16\kappa^2}\over\sqrt{1-16\kappa^2}} & 0 \\
	\sinh{1\over2}\gamma t & 0 & 0 & \cosh{1\over2}\gamma t 
	\end{pmatrix}\,.
	\end{equation}
\end{widetext} 
Hence, by taking the partial 
transpose of the associated Choi-Jamio\l kowski state, we can find the eigenvalues as functions of time:
\begin{equation}
	\lambda(t) = \begin{cases}
{1\over2}e^{-\gamma t/2}(\sinh(\gamma t/2)\pm 
Q_\kappa(\gamma t/2)), 
\\
{1\over2}e^{-\gamma t/2}(\cosh(\gamma t/2)\pm S_\kappa(\gamma t/2)),
	\end{cases}
\end{equation}
where in order to simplify the notation we have defined the functions
\begin{align}
	Q_\kappa(\tau) &= 
\sqrt{{\cosh^2(\tau\sqrt{1-16\kappa^2})-16\kappa^2}\over1-16\kappa^2}, \\
	S_\kappa(\tau) &= 
{\sinh(\tau\sqrt{1-16\kappa^2}))\over\sqrt{1-16\kappa^2}}.
\end{align}
By summing up the negative part of the eigenvalues, formula \eqref{NEGA} for 
the negativity of the Bit-Flip channel model follows.
In the same way, the Lindbladian \eqref{GAD_gen} is 
represented by the matrix \eqref{eq:GAD_matrix}.
By taking the exponential of \eqref{eq:GAD_matrix}, we have
\begin{widetext}
	\begin{equation}
	(2N+1)\Phi_t = \begin{pmatrix}
	N e^{-(2N+1)\gamma t}+N+1 & 0 & 0 & (N+1)(1-e^{-(2N+1)\gamma t}) \\
	0 & (2N+1)e^{-{1\over2}(2N+1)\gamma t} & 0 & 0 \\
	0 & 0 & (2N+1)e^{-{1\over2}(2N+1)\gamma t}  & 0 \\
	N(1-e^{-(2N+1)\gamma t}) & 0 & 0 & (N+1)e^{-(2N+1)\gamma t}+N \\
	\end{pmatrix}\;,
	\end{equation}
and therefore the associated Choi-Jamio\l kowski state reads
	\begin{equation}
	(4N+2)\rho_{AB}^{(\Phi_t)} = \begin{pmatrix}
	N e^{-(2N+1)\gamma t}+N+1 & 0 & 0 & (2N+1)e^{-{1\over2}(2N+1)\gamma t} \\
	0 & N(1-e^{-(2N+1)\gamma t}) & 0 & 0 \\
	0 & 0 & (N+1)(1-e^{-(2N+1)\gamma t})  & 0 \\
	(2N+1)e^{-{1\over2}(2N+1)\gamma t} & 0 & 0 & (N+1)e^{-(2N+1)\gamma t}+N \\
	\end{pmatrix}\;.
	\end{equation}
\end{widetext}
The eigenvalue equation for the partial tranpose of the Choi-Jamio\l kowski state 
yields
\begin{equation}
\lambda(t) = \begin{cases}
((N+1)e^{-(2N+1)\gamma t}+N)/(2N+1) \\
(N+1+Ne^{-(2N+1)\gamma t})/(2N+1) \\
{1\over2}e^{-(2N+1)\gamma t/2}(\sinh((2N+1)\gamma t/2)\pm 
A_N(\gamma t))
\end{cases}
\end{equation}
where we the function $A_N(\tau)$ was defined in \eqref{eq:amplitude}.
Summing up the negative parts of the eigenvalues, the formula \eqref{NENGA} 
for the negativity of the generalized amplitude channel model follows.

\section{Perturbative Expansion of EST}
\label{sec:perturb_th}
Let us re-write Eq.~\eqref{eq:maria} in terms of adimensional variables
\begin{equation}
\cosh(\tau) = 2+
{\cosh(\tau\sqrt{1-16\kappa^2})-16\kappa^2\over1-16\kappa^2
},
\label{eq:maria_2}
\end{equation}
where $\tau=\gamma t$ and $\kappa=\omega/\gamma$. We will now solve this 
equation in 
three different regimes.

By expanding the equation in powers of $\kappa$ and keeping 
only the first non-trivial term, we have the equation
\begin{equation}
 \tau\sinh(\tau)+2(1-\cosh(\tau)) \simeq {1\over4\kappa^2}.
\end{equation}
Furthermore we know that for small values of $\kappa$, $\tau_{ent}$ diverges because the 
process is 
asymptotically EB. Therefore, by expanding the equation for large $\tau$'s, we 
find
\begin{equation}
 \tau e^\tau \simeq {1\over2\kappa^2}, \quad\implies\quad {\cal T}_{ent}\simeq 
W\left({1\over2\kappa^2}\right).
\label{eq:bflip_eq_small}
\end{equation}
We can now find corrections perturbatively. Let us 
introduce the perturbative 
parmeter $\epsilon$ and consider the deformed equation
\begin{align}
 &\epsilon\cosh(\tau)+(1-\epsilon)\tau e^\tau = \nonumber \\ 
 &2\epsilon+
\epsilon{\cosh(\tau\sqrt{1-16\kappa^2})-16\kappa^2\over1-16\kappa^2
} + {(1-\epsilon)\over2\kappa^2}.
\label{eq:deformed_1}
\end{align}
This equation interpolates between the known asymptotic value of ${\cal 
T}_{ent}$ 
at 
$\epsilon=0$ and the unknown value of ${\cal T}_{ent}$ at $\epsilon=1$. We 
therefore look 
for 
solutions to the above equation in the form
\begin{equation}
 \tau(\epsilon) = \sum_{n=0}^\infty \tau_n(\kappa)\epsilon^n.
 \label{eq:perturbative_expansion}
\end{equation}
If the series converges, we have ${\cal T}_{ent}(\kappa) = \sum_{n=0}^\infty 
\tau_n(\kappa)$. By 
expanding Eq.~\eqref{eq:deformed_1} in powers of $\epsilon$ and applying 
Eq.~\eqref{eq:perturbative_expansion} to it, we can recursively determine the 
coefficients $\tau_n(\kappa)$ by imposing equality order-by-order in 
$\epsilon$. The 
first 
correction to~\eqref{eq:bflip_eq_small} turns out to be
\begin{align}
 \tau_1(\kappa) &= 
{
(1-2\kappa^2W(1/2\kappa^2))^2-4\kappa^2W(1/2\kappa^2)
\over2(1+W(1/2\kappa^2 ) ) }
\nonumber \\
&+
{1+(2\kappa^2W(1/2\kappa^2))^{1-16\kappa^2}
\over2\sqrt{1-16\kappa^2}(2\kappa^2W(1/2\kappa^2))^{\sqrt{1-16\kappa^2}-1}}.
\end{align}

At the critical ratio $\tau=1/4$ Eq.~\eqref{eq:maria_2} simplifies 
considerably, 
leaving us with
\begin{equation}
 \cosh(\tau)-1-{\tau^2\over2} = 2 \quad\implies\quad {\cal T}_{ent}\simeq2.5.
\end{equation}
We can expand around this value by looking for solutions of the form
\begin{equation}
 {\cal T}_{ent}(\kappa) = \sum_{n=0}^\infty 
\tau_n\left(\kappa-{1\over4}\right)^n.
\end{equation}
We can determine the coefficients $\tau_n$ recursively by plugging the above 
expansion into Eq.~\eqref{eq:maria_2} and expanding it in power of 
$(\tau-1/4)$. 
The first few coefficients of the expansion read
\begin{align}
 \tau_0 &\simeq 2.5, \\
 \tau_1 &= -{\tau_0^4\over3(\sinh(\tau_0)-\tau_0)} \simeq -3.7, \\
 \tau_2 &= 
{
\tau_0^4(4\tau_0^2\cosh^2(\tau_0)-5\tau_0^4\cosh(\tau_0)) 
\over90(\sinh(\tau_0)
 - \tau_0)^3} \nonumber \\
&+{
4\tau_0^4((\tau_0^2-15)\sinh^2(\tau_0)+6\tau_0(\tau_0^2+5)\sinh(\tau_0))\over9
0 (\sinh(\tau_0) - \tau_0)^3} \nonumber \\
&+{\tau_0^2(27\tau_0^2+64))\over9
0 (\sinh(\tau_0) - \tau_0)^3}\simeq 10.6\,.
\end{align}

By taking the limit $\kappa\rightarrow+\infty$ of Eq.~\eqref{eq:maria_2}, we 
have 
the simple solution
\begin{equation}
\lim_{\kappa\rightarrow\infty} {\cal T}_{ent}(\kappa) = 
\mathrm{arcosh}(3): = {\cal T}_{ent}(\infty).
\end{equation}
One might therefore think of looking for corrections by
expanding ${\cal T}_{ent}$ as a Laurent series:
${\cal T}_{ent}(\kappa) = \sum_{n=0}^{\infty}{\tau_{2n}\over \kappa^{2n}}$.
However, when $\kappa$ is continued to the 
complex plane, Eq.~\eqref{eq:maria_2} has an essential singularity at 
$\kappa=\infty$. As a result, an expansion of the form above cannot be found. 
Instead, we can obtain a 
perturbative series about infinity by introducing a perturbative parameter 
$\epsilon$, deforming Eq.~\eqref{eq:maria_2} into
\begin{equation}
\cosh(\tau) = 2+
\epsilon\,{\cosh(\tau\sqrt{1-4\kappa^2}
)-4\kappa^2\over1-4\kappa^2
} + 1-\epsilon.
\label{eq:deformed_maria}
\end{equation}
We can thus perform a perturbative expansion similar to the $
\kappa\simeq0$ regime: 
\begin{equation}
\tau(\epsilon) = \sum_{n=0}^{\infty}\tau_n(\kappa)\epsilon^n,
\end{equation}
the first few coefficients of the expansion being 
\begin{align}
 \tau_0(\kappa) &= {\cal T}_{ent}(\infty), \\
 \tau_1(\kappa) &= {\cosh({\cal T}_{ent}(\infty)\sqrt{1-16\kappa^2})-1\over 
2\sqrt{2}(1-16\kappa^2)}.
\end{align}

\section{Intro to Gaussian Bosonic Channels} \label{Sec:GAUSAPPENDIX} 
A formal definition of Gaussian Bosonic Channels can be obtained by passing into the Heisenberg representation~\cite{HOLEREV} and assigning their action on the Weyl operators of the system~\cite{holevo07}. 
We remind that assuming the system of interest to be composed by $n$ independent modes described by 
canonical coordinates $\{ Q_j,P_j\}_{j=1,\cdots,n}$ fulfilling the 
 the canonical commutation relations
\begin{equation}
 [Q_i,P_j] = i\,\delta_{ij}, \qquad [Q_i,Q_j] = [P_i,P_j] = 0,
\end{equation}
a generic Weyl operator  is defined as the unitary 
transformation $W_\xi=e^{i\xi\cdot R}$  with 
with $\xi\in\R^{2n}$ and  $R=(Q_1,P_1,...,Q_n,P_n)$.
 A zero-mean Gaussian channel $\Phi$ can then be uniquely identified by 
 two $2n\times 2n$ real matrices $F$ and $G$ that, in the Heisenberg representation,  define 
the mapping 
\begin{equation}
W_\xi\quad  \longrightarrow  \quad  W_{F\xi}\,e^{-{i\over2}\xi^T G\xi}, \qquad \forall \xi\;. 
\end{equation}
The CPt condition imposes on $F,G$ the 
inequality
\begin{equation}
 G \geq {i\over2}(J-F^TJF),
 \label{eq:cpt_gaussian_condition}
\end{equation}
where $J$ is the {standard symplectic metric} of the system, i.e. 
\begin{equation}
 J = \bigoplus_{i=1}^n\begin{pmatrix}
  0 & 1 \\
  -1 & 0
 \end{pmatrix}.
\end{equation}
It can be proven~\cite{holevo08} that a Gaussian channel $(F,G)$ is EB if and 
only if it admits a decomposition of the form $G=\mi+\nee$, and such that
\begin{equation}
	\nee\geq{i\over2}\,J, \qquad \mi\geq{i\over2}\,F^TJF. 
	\label{eq:gaussian_eb_condition}
\end{equation}
Therefore a necessary condition for $(F,G)$ to be EB is
\begin{equation}
  G \geq {i\over2}(J+F^TJF).
  \label{eq:gaussian_nec_eb}
\end{equation}

Let us now restrict our attention to one-mode Gaussian channels (i.e. $n=1$). 
For such channels, a complete characterization 
can be given, based upon the Williamson theorem~\cite{holevo07,holevo01}. As it 
turns out, depending on the value of the quantity $F^TJF$, there exists 
canonical unitary transformations $U_1,U_2$ such that, via the mapping
\begin{equation}
	\Phi(\rho) \longrightarrow U_2\,\Phi(U_1\,\rho\, 
U_1^\dagger)\,U_2^\dagger,
\end{equation}
the Gaussian channel $\Phi\simeq(F,G)$ can be reduced to one of 
the following 
normal forms
\begin{enumerate}
	\item[$A$)] $F^TJF=0$. Then $(F,G)$ can be reduced to the form
	\begin{equation}
		F= k\,E^{(00)}, \qquad G = \left(q+{1\over2}\right)I.
	\end{equation}
	\item[$B_1$)] $F^TJF=J$. Then $(F,G)$ can be reduced to the form
	\begin{equation}
	F = I, \qquad G = {1\over2}\,E^{(11)}.
	\end{equation}
	\item[$B_2$)] $F^TJF=J$. Then $(F,G)$ can be reduced to the form
	\begin{equation}
	F = I, \qquad G = q\,I.
	\end{equation}
	\item[$C$)] $F^TJF=k^2J$, $k>0,k\neq1$. Then $(F,G)$ can be reduced to 
the form
	\begin{equation}
	F = k\,I, \qquad G = \left(q+{|1-k^2|\over2}\right)I.
	\end{equation}
	\item[$D$)] $F^TJF=-k^2J$, $k>0$. Then $(F,G)$ can be reduced to the 
form
	\begin{equation}
	F = k\,Z, \qquad G = \left(q+{1+k^2\over2}\right)I.
	\end{equation}
\end{enumerate}
where $k$ is a real number and $q\geq0$. Combining the above result 
with~\eqref{eq:gaussian_eb_condition}, we have the following EB 
conditions for one-mode Gaussian channels~\cite{holevo08}:
\begin{enumerate}
	\item[$A)$] $\Phi$ is EB (in fact, it is c-q).
	\item[$B_1)$] $\Phi$ is not EB.
	\item[$B_2)$] $\Phi$ is EB if and only if $q\geq1$.
	\item[$C)$] $\Phi$ is EB if and only if $q\geq\min\{1,k^2\}$.
	\item[$D)$] $\Phi$ is EB.
\end{enumerate}

Notice that the only channels which could either be EB or non-EB, depending on $k$ and $q$, are the 
ones 
in the classes $B_2$ and $C$, and that for 
{them  Eq.~\eqref{eq:gaussian_nec_eb} is in fact equivalent to the EB 
conditions~(\ref{eq:gaussian_eb_condition}). Furthermore, by continuity,  for these maps the entanglement transmission time can be 
determined studying the zeros of the following equation}  \begin{equation} \label{IMPO1} 
 \det\left(G-{i\over2}(J+F^TJF)\right) = 0.
\end{equation} 

\section{EST for the Gaussian amplitude damping channel} \label{APPFIN} 

For the model described by the GKLS generator in Eq.~\eqref{LINDG} and~\eqref{HAMG}, the matrices $F_t$ and $G_t$ can be espressed as
\begin{align}
	F_t &= \exp\left[\gamma t\begin{pmatrix}
	-{1\over2}N+\kappa\sin\theta & -\kappa\cos\theta \\
	\kappa\cos\theta & -{1\over2}N-\kappa\sin\theta
	\end{pmatrix}\right], \\
	G_t &= (2N+1)\gamma\int_{0}^{t}\diff s\,F_s^TF_s.
\end{align} 
More explicitly, the matrix elements of $F_t$ are
\begin{align}
	F_t|_{11} &= {i\over2}e^{-{1\over2}N\tau}\Bigg[\left({\sin\theta\over\sqrt{\cos2\theta}}-i\right)e^{-i\kappa\tau\sqrt{\cos2\theta}} \nonumber\\ &\qquad\qquad\quad -\left({\sin\theta\over\sqrt{\cos2\theta}}+i\right)e^{i\kappa\tau\sqrt{\cos2\theta}}\Bigg] \\
	F_t|_{21} &= {\cos\theta\over\sqrt{\cos2\theta}}e^{-{1\over2}N\tau}\sin(\kappa\tau\sqrt{\cos2\theta}) \\
	F_t|_{12} &= -F_t|_{21} \\
	F_t|_{22} &= -{i\over2}e^{-{1\over2}N\tau}\Bigg[\left({\sin\theta\over\sqrt{\cos2\theta}}+i\right)e^{-i\kappa\tau\sqrt{\cos2\theta}} \nonumber\\ &\qquad\qquad\quad-\left({\sin\theta\over\sqrt{\cos2\theta}}-i\right)e^{i\kappa\tau\sqrt{\cos2\theta}}\Bigg], \\
\end{align}
where $\tau=\gamma t$, while for the matrix $G_t$ we have
\begin{align}
G_t|_{11} &= -{(2N+1)\sec2\theta\over1+4\kappa^2\cos2\theta}\Bigg[e^{-\tau}\kappa^2-\cos2\theta(1+2\kappa^2 \nonumber\\
&+2\kappa(\kappa\cos2\theta+\sin\theta))+ {1\over2}e^{-\tau}(2\kappa^2\cos4\theta \nonumber\\
&+\cos2\theta(1+4\kappa^2+\cos(2\kappa\tau\sqrt{\cos2\theta}))(1+4\kappa\sin\theta) \nonumber\\
&+2\sqrt{\cos2\theta}\sin\theta(2\kappa\sin\theta+1)\sin(2\kappa\tau\sqrt{\cos2\theta})) \nonumber \\
&\qquad\qquad\qquad\qquad\qquad+2\sin^2(\kappa\tau\sqrt{\cos2\theta})\Bigg] \\
G_t|_{21} &= {(2N+1)\tan2\theta\over1+4\kappa^2\cos2\theta}e^{-\tau}\Bigg[2(1-e^\tau)\kappa^2\cos2\theta \nonumber\\
&+\kappa\sqrt{\cos2\theta}\sin(\kappa\tau\sqrt{\cos2\theta}) + \sin^2(\kappa\tau\sqrt{\cos2\theta})\Bigg]\\
G_t|_{12} &= G_t|_{21} \\
G_t|_{22} &= -{(2N+1)\sec2\theta\over1+4\kappa^2\cos2\theta}\Bigg[e^{-\tau}\kappa^2-\cos2\theta(1+2\kappa^2 \nonumber\\
&+2\kappa(\kappa\cos2\theta-\sin\theta))+ {1\over2}e^{-\tau}(2\kappa^2\cos4\theta \nonumber\\
&+\cos2\theta(1+4\kappa^2+\cos(2\kappa\tau\sqrt{\cos2\theta}))(1-4\kappa\sin\theta) \nonumber\\
&+2\sqrt{\cos2\theta}\sin\theta(2\kappa\sin\theta-1)\sin(2\kappa\tau\sqrt{\cos2\theta})) \nonumber \\
&\qquad\qquad\qquad\qquad\qquad+2\sin^2(\kappa\tau\sqrt{\cos2\theta})\Bigg].
\end{align}
Let us now set $\gamma_1=\gamma(N+1)$ and $\gamma_2=\gamma N$ for the sake of conveniency. Eq.~(\ref{IMPO111}) gives the following equation for the EST 
\begin{equation}
 \cosh(\tau) -a(\gamma_1,\gamma_2,\kappa,\theta) 
\cos(2\tau\kappa\sqrt{\cos2\theta}) = 
b(\gamma_1,\gamma_2,\kappa,\theta) \label{QUESTAQUI} 
\end{equation}
where 
\begin{align}
 a &= 
{
2(\gamma_1+\gamma_2)^2(1-\sec2\theta)\over(3\gamma_1+\gamma
_2) (\gamma_1 + 3\gamma_2) + 4\kappa^2\beta},  \\
 b &= 
{
2(\gamma_1+\gamma_2)^2\sec2\theta 
+ (3\gamma_1^2+2\gamma_1 \gamma_2 +
3\gamma_2^2) + 4\kappa^2\alpha
 \over(3\gamma_1+
\gamma_2) (\gamma_1 + 3\gamma_2) + 4\kappa^2\beta} , \\
\alpha &= ( 
\gamma_1^2+\gamma_2^2)(2+3\cos2\theta)+2\gamma_1\gamma_2(2+\cos2\theta), \\
\beta &= ( 
\gamma_1^2+\gamma_2^2)(2+\cos2\theta)+2\gamma_1\gamma_2(2+3\cos2\theta).
\end{align}
The equation presents a critical behavior at $\theta=\pi/4$, and becomes
\begin{align} 
&\cosh(\tau) = \nonumber \\
&{
(\gamma_1-\gamma_2)^2+4(\gamma_1+\gamma_2)^2+4\kappa^2(\gamma_1+\gamma_2)^2(2+\tau
^2)
 \over 
(3\gamma_1+\gamma_2)(\gamma_1+3\gamma_2)+8\kappa^2(\gamma_1
+ \gamma_ 2)^2},
\end{align}
while for $\kappa=0$ it yields  the purely 
dissipative EST value 
\begin{equation}
 {\cal T}_{ent} = \ln{3\gamma_1+\gamma_1\over\gamma_1+3\gamma_2},
\end{equation}
we have reported as  Eq.~(\ref{TAUTAUE}) of   the main text.

{For small values of $\kappa$   we  look for solutions of the form
\begin{equation}
 {\cal T}_{ent}(\kappa)=\sum_{n=0}^\infty \tau_n\kappa^n.
\end{equation}
The first corrections are 
\begin{align}
 \tau_0 &= \ln{3\gamma_1+\gamma_2\over\gamma_1+3\gamma_2}\;, \qquad 
\qquad  \tau_1 = 0\;, \\
 \tau_2 &= 
{\gamma_1+\gamma_2\over\gamma_1-\gamma_2}\Bigg[{
4(\gamma_1-\gamma_2)^2\over(3\gamma_1+\gamma_2)(\gamma_1+3\gamma_2)} \nonumber 
\\ 
&\qquad\qquad\quad-\ln^2{3\gamma_1+\gamma_2\over\gamma_1+3\gamma_2}
\Bigg](1-\cos2\theta)\;, 
\end{align}
and this is valid for all forms of the equation.

Let us now consider the behaviour of the equation for large $\kappa$. When 
$0\leq\theta<\pi/4$, the by taking the limit we have
\begin{equation}
 \cosh(\tau) = \lim_{\kappa\rightarrow\infty}b(\gamma_1,\gamma_2,\kappa,\theta)
\end{equation}
and therefore
\begin{align}
 \lim_{\kappa\rightarrow+\infty}{\cal T}_{ent}(\kappa) = {\rm 
arcosh}{\alpha(\gamma_1,\gamma_2,\theta)\over\beta(\gamma_1,\gamma_2,\theta)}.
\end{align}
Using the same deformation procedure employed for Eq.~\eqref{eq:deformed_maria} we can find the first correction
\begin{equation}
 {a\cos(2\kappa\sqrt{-\cos2\theta}\tau_0)+b-b_\infty\over\sqrt{b_\infty^2-1}}\;,
\end{equation}
where $b_\infty=\lim\limits_{\kappa\rightarrow\infty}b(\kappa)$, yielding Eq.~(\ref{FFFDD}) of the main text. Instead for 
$\theta\in(\pi/4,\pi/2)$, we have asymptotically for 
$\kappa\rightarrow+\infty$
\begin{equation}
 {\cal T}_{ent} \simeq {1\over2\kappa\sqrt{-\cos2\theta}}{\rm 
arcosh}\,{2\kappa^2\alpha(\gamma_1,\gamma_2,\theta)
 \over 
(\gamma_1+\gamma_2)^2(\sec2\theta-1)}\;,
\end{equation}
that coincides with Eq.~(\ref{ENEW}) 
of the main text. }

\end{document}